# Spatial Division Multiplexing for Multiplex Coherent Anti-Stokes Raman Scattering

T. Mansuryan, A. Tonello, K. Krupa, A. Desmouliére, G. Ndong Ntoutoume, V. Sol, C. Lefort, M. Zitelli, M. Ferraro, F. Mangini, Y. Sun, Y. Arosa Lobato, B. Wetzel, S. Wabnitz *Fellow, IEEE*, and V. Couderc

*Abstract*— We demonstrate how a narrowband pump and a broadband spectrum can be spatially multiplexed by selective coupling them in two distinct modes of a few-mode microstructure fiber. The first mode carries most of the input pump energy, and experiences spectral broadening. Whereas the second mode preserves the narrow bandwidth of the remaining part of the pump. Bimodal propagation, with a power unbalance strongly in favor of the fundamental mode, is naturally obtained by maximizing coupling into the fundamental mode of the fiber. At the fiber output, the nearly monochromatic beam and the supercontinuum carried by the two different modes are combined by a microscope objective, and used as a pump and a Stokes wave for self-referenced multiplex coherent anti-Stokes Raman scattering micro-spectroscopy. The spectral resolution, the signal-to-noise-ratio, and the possible amplification of the remaining pump beam are discussed.

## I. INTRODUCTION

FOLLOWING the work of Alfano and Shapiro [1]-[2] on supercontinuum generation (SCG), the advent of microstructured optical fibers (MOF) [3]-[5] opened a new avenue for the generation of polychromatic beams [6]. MOF provide tight confinement of light beams, and fiber dispersion can be tailored by changing the fiber transverse structure, permitting to control and selectively enhance nonlinear propagation effects [7], [8]. SCG covering the entire transparency window of silica glass was reported [6]-[9]; by suitable waveguide design and material selection, SCG may extend up to the mid-infrared [10]. Based on either femtosecond, picosecond or nanosecond laser pumps, MOF-based SCG found successful applications in many fields, and especially in multispectral imaging systems. Although using supercontinuum lasers for biomedical applications was first reported back in the nineties [9], the advent of MOF has permitted to significantly increase the SCG bandwidth and spectral power density [12], [13], with a significant impact on imaging system improvements.

The first demonstrations of SCG in MOF were performed by using femtosecond pulses and operating with a laser carrier wavelength close to the fiber zero-dispersion wavelength (ZDW) [9]. At present days, the principal mechanisms of SCG are well understood, and they involve a variety of nonlinear effects such as: self-phase modulation, cross-phase modulation, four-wave mixing, soliton propagation, soliton fission and stimulated Raman scattering [9], [14], [15]. However, when using several meters of fiber, fiber dispersion separates wave-packets at different central wavelengths. This fact can limit the SCG applicability, whenever synchronous wave-packets are instead required. Although long pulses are less affected by fiber dispersion, in regimes involving high peak powers and pulse energies, there is an intrinsic limit of SCG applicability for illuminating samples, such as biomedical living targets, having a low damage threshold.

Multiplex Coherent anti-Stokes Raman scattering (M-CARS) is a nonlinear spectroscopic technique developed to identify the vibrational modes, hence the molecular structures of samples [16], [17]. This technique requires the synchronous nonlinear interplay between a narrowband pump and a broadband Stokes spectrum at the sample level, and permits the simultaneous detection of all vibrational modes which are covered by pump-Stokes beating. As discussed below, such a technique often requires the use of sub-nanosecond pulses, in order to keep all supercontinuum (SC) wavelengths temporally superimposed [18], [19]. Nonlinear spectral broadening of long pulses in MOF causes large depletion and modulation of the pump envelope in the time domain, thus hampering the synchronization between pump end Stokes wavelengths. In turn, this effect prohibits the direct exploitability of both residual pump and SC for M-CARS. To overcome this

This work was supported in part by Arianegroup (X-LAS); Conseil Régional Aquitaine (F2MH, FLOWA, SI2P, SPINAL); Agence Nationale de la Recherche (ANR-10-LABX-0074-01, ANR-18-CE080016-01); European Research Council (740355, 950618); Ministero dell'Istruzione, dell'Università e della Ricerca (R18SPB8227); Polish National Agency for Academic Exchange and French-Polish Partnership Hubert Curien (POLONIUM program N.BPN/BFR/2021/1/00013, 48161TH).

T. M., A. T., C. L., B. W. and V. C. are with Limoges University, XLIM, UMR CNRS 7252, 123 Av. A. Thomas, 87060 Limoges, France. (V. C. email address is vincent.couderc@xlim.fr).
K. K. is with Institute of Physical Chemistry, Polish Academy of Sciences, Warsaw, ul. Kasprzaka 44/52, 01-224 Poland.
M. Z., M. F., Y. S., and S. W. are with Dipartimento di Ingegneria dell'Informazione, Elettronica e Telecomunicazioni, Sapienza University of Rome, Via Eudossiana 18, 00184 Rome, Italy.
S. W. is also with CNR-INO, Istituto Nazionale di Ottica, Via Campi Flegrei 34, 80078 Pozzuoli, Italy
F. M. is with University of Brescia, Department of Information Engineering (DII), via Branze 38, 25123, Brescia, Italy.
G. N. N. and V. S. are with Limoges University, LABCiS UR 22722, F-87000 Limoges, France
A. D. is with Université de Limoges, UR 20218 NeurIT, Faculties of Medicine and Pharmacy, F-87000 Limoges, France
Y. A. L. is with University of Santiago de Compostela Praza do Obradoiro S/N, Santiago de Compostela (15782) Coruña, (SPAIN).

temporal mismatch problem, often in M-CARS set-ups a portion of the input narrow-band pump is collected before injection in the fiber, and propagated through a delay line to ensure its temporal overlap with the fiber SC. In addition, the pump and the SC are spatially superimposed, and sent to the sample under test. In the work of Mikami [20], a compact source for CARS microscopy was demonstrated, using nanosecond pulses from a microchip laser at 1064 nm for pumping a short length of MOF in the anomalous dispersion regime. They claim that short MOF length permitted to maintain temporal overlap between the residual pump pulse and the SC, however there is no detailed numerical or experimental analysis provided. And also CARS signal intensity was limited by the damage threshold of the MOF. Recently, S. Wehbi *et al.* have shown how a self-referenced M-CARS system can be obtained from the spatiotemporal Kerr-induced self-cleaning process in multimode graded-index fibers [21], [22]. In that experiment, the residual pump was shown to be partially recovered at the fiber output, as a result of the energy exchange among a multitude of fiber modes along their propagation. These approaches permit a significant simplification of the M-CARS setup, making it free of the external delay line. Besides the technical problems related to the presence of the delay line, it is relevant to recall that CARS spectroscopy has some intrinsic limitations, which requires signal post-processing: in particular, the output signal of this nonlinear vibrational spectroscopy technique is partially affected by a non-resonant background, which limits its signal-to-noise ratio [23].

In this paper, we show that using a few-mode MOF permits to combine the two wave-packets (pump and SC) in a single waveguide. Although pump laser coupling was optimized in order to maximize the power on the MOF fundamental mode ($LP_{01}$), it is easy to detect at the fiber output the presence of a guided first odd-parity mode ($LP_{11}$), carrying a minor fraction of the guided power. Such effective unbalanced multimode coupling of a pump beam in a few-mode MOF enables efficient spectral broadening in one mode, while leaving the residual pump that propagates in the other mode nearly unaffected, and yet temporally synchronized. Specifically, we show that a second guided mode can be judiciously used as a second channel, thus assuring both spatial and temporal overlap between the pump and the broadband Stokes waves when light is focalized on the sample. Nonlinear spectral broadening is a powerful tool to generate a broad spectrum: however, any spectral broadening in the residual pump will degrade the spectral resolution of the M-CARS signal. In this work, we show how the residual pump in the few-mode MOF is only weakly affected by nonlinear spectral broadening, in spite of its co-propagation with the fundamental mode pumping the SCG. The few-mode MOF allows to carry in parallel both the nonlinear spectral broadening of the field in the fundamental mode, and the residual pump field in the first high-order mode, despite inherent cross-phase modulation effects. As a proof of principle, the resulting beam is used for broadband excitation and an M-CARS implementation on a paraffin sample, as a simple proof-of-principle M-CARS spectral measurement, and on a curcumin superparticle [24], [25], as multimodal imaging example.

## I. DESCRIPTION OF THE EXPERIMENT

The schematic of experimental setup is shown in Fig. 1. We use a micro-chip laser, delivering 1 ns pulses at 1064 nm, with a repetition rate of 20 kHz and a Gaussian beam shape. The input power is coupled into a few-mode MOF by means of an aspherical lens (AL) with 4 mm focal length. Optimization of laser coupling into the MOF inherently leads a small fraction of the input power (estimated at 5%) to be coupled in the first odd-parity mode ($LP_{11}$), whereas the main part of the input (95%) is carried by the fundamental mode. The input polarization state is linear, and can be further rotated by means of a half-wave plate. As a result, two different spatial modes co-propagate in the MOF (not considering polarization degeneracy). We used a 2.8 m long air-silica MOF, whose transverse structure is illustrated in Fig. 1(a). The fiber core has a diameter of 4 μm, and it is surrounded by a hexagonal lattice with air-hole diameter of 3 μm and spacing of 4 μm. The dispersion curves of the fiber $LP_{01}$ and $LP_{11}$ modes, with a zero-dispersion wavelength of 980 nm and 850 nm respectively, are reported in Fig. 1(b).

Output beam from MOF is collimated by parabolic mirror (PM) to avoid chromatic distortions. Next, we use a long-pass filter (F1, Thorlabs FELH1000), in order to suppress any wavelengths below 1000 nm. The filtered beam is then focalized on the sample with a large numerical aperture (NA=1.1) objective (Olympus LUMFLN60XW). The resulting M-CARS signal, ranging from 4000 $cm^{-1}$ to 0 $cm^{-1}$, is collected by another 60x objective, then filtered out by s short-pass filter (F1, Thorlabs FELH1000) in order to keep only CARS signal, and finally analyzed by the spectroscope (LabRAM HR Evolution from Horiba). The sample is fixed on a 2D automated translation stage, and its mechanical scanning finally permits to collect the complete CARS spectrum pixel by pixel, and to reconstruct the M-CARS images.

## II. SUPERCONTINUUM GENERATION

As already mentioned, in our experiment the 1064 nm laser beam was coupled into both $LP_{01}$ and $LP_{11}$ modes of the MOF with an initial power ratio of 95% and 5%, respectively. Pump light propagation in the anomalous dispersion regime is initially affected by modulation instability. As a result, the nanosecond input pulse breaks up in many ultrashort pulses which, in turn, evolve into optical solitons. As well known, Raman gain provides a natural mechanism of soliton self-frequency shift, causing the spectrum to expand towards the infrared spectral domain. At the same time, soliton propagation induces dispersive wave emission in the visible range of the spectrum (propagating in the normal dispersion regime); the efficiency of this process is enhanced when solitons and dispersive waves are group-velocity matched. As a result, a SCG ranging from 0.5 μm to 2.4 μm is obtained (Fig. 2(a)). In Fig. 2(b) we can follow the spectral broadening evolution versus injection power. Due to the dominant contribution of the $LP_{01}$ mode, all of the converted wavelengths are carried by the same fundamental mode (see Fig. 2(b) insets).

The process of SCG, and in particular Raman soliton self-frequency shift, strongly depletes the central part of the temporal envelope of the field carried by mode $LP_{01}$, leaving unaffected only the leading and trailing edges of the residual pulse at 1064 nm (Fig. 2(c)). In Fig. 2(d) we can follow the distortion evolution of the time profile of the pump wavelength as a function of the injection power. Such a pulse distortion severely reduces any possibility of nonlinear interplay between the residual pump and the generated SC, owing to their relative temporal desynchronization. Thus, when considering only SCG in the $LP_{01}$ mode, M-CARS cannot be observed in our experiments. However, when a small part of the pump power is carried in parallel by the mode $LP_{11}$, it is possible to retain a certain amount of unconverted input pulse energy at 1064 nm within the central part of the pump pulse, which is still synchronized with the SC. Indeed, due to poor input beam coupling in the $LP_{11}$ mode, no significant spectral broadening was observed for light carried in this mode, despite the presence of cross-phase modulation induced by the $LP_{01}$ mode (considering that the Stokes spectrum can host high peak power solitons). In this case, the portion of the pump beam at 1064 nm carried by the $LP_{11}$ mode can readily interact with the infrared SC, thus generating the desired M-CARS signal, without the need for an additional delay line or any further pump-probe beam synchronization.

### III. SPATIOTEMPORAL DISTRIBUTION OF THE BEAM ENERGY

In order to provide a better understanding of the physical mechanisms behind our self-referenced M-CARS approach, we analyzed the energy distribution of the beam at the fiber output in the temporal, spatial and spectral domains. Spectral evolution of SC and temporal evolution of pump pulse are discussed in previous section.

Here we go further by exploring spectro-temporal evolution of SC and spatio-temporal evolution of pump pulse. Fig. 3(a) illustrates spectro-temporal energy repartition, or, in other words, spectral decomposition of temporal profile of output pulse. In Fig. 3(b) we can see some selected wavelengths temporal profile, including low power pump wavelength for comparison. Let us recall that, for each pulse, power is distributed over the $LP_{01}$ (95%) and the $LP_{11}$ (5%) modes. At low input peak power (0.25 kW), no significant spectral broadening nor temporal reshaping is observed for the pump wavelength (Fig. 3(b), lowest trace). The pulse preserves its nearly-Gaussian, slightly asymmetrical shape (the slight pulse asymmetry is caused by the passive Q-switching process inherent to the laser source. The corresponding output spatial beam exhibits a nearly Gaussian profile (Fig. 3(c), left profile). However, when increasing the input peak power up to 5 kW, the output spectrum starts to significantly broaden, and a SC is progressively generated. The output pulse at the pump wavelength (1064 nm) is affected by significant depletion and distortions, with the presence of multiple peaks in its trailing edge (Fig. 3 (a) and 3(b), trace #0). However, the pump fraction carried by the $LP_{11}$ mode is conserved, and it remains synchronized with all the wavelengths of the SC (Fig. 3(a) and 3(b), temporal interval between dashed lines). Few-modal nature of propagation is obvious at high power (central profile in Fig. 3(d)), especially when decomposed by polarization (Fig. 3(d), upper and lower profiles). Finally we studied temporal profiles of various part of this spatial profile (Fig.3 (e)). We can see that the central part of the beam is completely depleted, while the peripheries still carries some remaining energy (Fig. 3(e), temporal interval between dashed lines).

SC is mainly generated in the central part of the beam, but there is some recovering with the synchronous pump of periphery, especially when strongly focused (chromatic aberration) on sample.

These spatial and temporal synchronization of pump and Stokes wavelengths assuring our self-referenced M-CARS experiment.

### IV. NUMERICAL SIMULATIONS

A numerical model suitable to analyze our experiments can be based on a pulse propagation of two generalized nonlinear Schrödinger equations (GNLEs), one per mode, in the presence of dispersion, instantaneous and delayed nonlinearity, and difference in mode group velocities. Nonlinear coupling between the two GNLEs was limited to the sole contribution of cross-phase modulation, owing to the large difference in propagation constants. The two GNLEs are described in the supplemental information. Their basic structure, referred to Mode 0 ($LP_{01}$) and Mode 1 ($LP_{11}$) is as follow:

$$\left[\frac{\partial}{\partial z} - i\widehat{D}_0\right]A_0 = i\widehat{N}\{(1-f_R)A_0[S_{0000}|A_0|^2 + 2S_{0011}|A_1|^2] + f_R R_0(t,z)\}$$

$$\left[\frac{\partial}{\partial z} - i\widehat{D}_1\right]A_1 = i\widehat{N}\{(1-f_R)A_1[S_{1111}|A_1|^2 + 2S_{1010}|A_0|^2] + f_R R_1(t,z)\} \quad (1),$$

where the operator $\widehat{N} = n_2\omega_0(1 + i\tau_S \partial/\partial t)/c$, and $R_h(t,z)$ accounts for the Raman effect on mode h, h being 0 or 1.

The linear propagation operators $\widehat{D}_h$ and mode profiles have been calculated by means of a mode solver. Eqs. (1) include both the instantaneous Kerr effect and the delayed Raman nonlinear response. We neglected various other nonlinear terms of modal interplay, due to the sufficiently large difference in propagation constants, and the expected poor spatial overlap among the considered modes. In the simulations, the parameters used for the Raman response

$$h_R(t) = \frac{\tau_1^2 + \tau_2^2}{\tau_1 \tau_2^2} e^{-\frac{t}{\tau_2}} sin\left(\frac{t}{\tau_1}\right) \quad (2)$$

are typical for silica with a Raman fraction $f_R=0.18$ and time constants $\tau_1=12.2$fs, and $\tau_2=32$fs. We also included frequency-dependent linear losses to model the strong glass absorption above 2400nm.

At the pump wavelength, the calculated relative delay between mode $LP_{11}$ and mode $LP_{01}$ is of 35ps/m. The calculated overlap integrals are $A_{eff} = 1/S_{0000} = 10.7\mu m^2$, $S_{1111} = 1.06 S_{0000}$, $S_{1010} = S_{0011} = 0.66 S_{0000}$. For the input conditions, we consider a Gaussian pulse of 2 kW peak power with a duration of 320 ps. The pulse duration was reduced with respect to the experiments, to alleviate the computation time of our simulations (as the dynamics can be considered quasi-continuous and essentially peak power-dependent in the long nanosecond pulse regime). As observed in the experiment, the considered input pulse is not symmetrical; we use a fundamental mode carrying 96% of the input power, while the remaining part is carried by the $LP_{11}$ mode.

Fig. 4(a) illustrates the total output spectrum. The calculated spectro-temporal distributions, obtained on both $LP_{01}$ and $LP_{11}$ modes, are shown in Fig. 4(b) and (c). As expected, the main spectral broadening is observed on the fundamental mode ($LP_{01}$), with a spectral extension up to 2.4 µm. We see that, in the fundamental mode, the temporal spreading of the output pulse(s) gradually decreases when increasing the wavelength. In agreement with experiments, in the $LP_{01}$ mode the residual pump at 1064 nm is strongly depleted in the time window of the supercontinuum generation. In contrast, when considering the $LP_{11}$ mode, which is seeded at much lower power and having a small overlap integral with the fundamental mode, its energy remains confined near the pump wavelength, without frequency conversion. Under these conditions, the spectral broadening of the pump beam propagating on $LP_{11}$ remains fairly limited. One can also observe in Fig. 4(c) the high temporal localization of energy at 1064 nm within the $LP_{11}$ mode, and its temporal synchronization with other selected supercontinuum wavelengths at 1080 nm, 1200 nm and 1500 nm (within the $LP_{01}$ mode), which in turn will favor the excitation of the highly sought-after M-CARS signatures (see Fig. 4(b)). Panel (d) of Figure 4 illustrates the output pulse envelopes at 1064 nm, after a bandpass filter with 25GHz bandwidth: while the temporal envelope of the $LP_{01}$ mode is strongly depleted around the pulse center, the pump wavelength remains carried by the weak pulse envelope of the $LP_{11}$ mode at all times.

## V. M-CARS Experiment

For our proof-of-concept demonstration, the M-CARS system exploiting few-mode coupling within the MOF was first tested with a liquid paraffin sample. In Fig. 5 we compare the M-CARS spectra obtained with our self-referenced setup and those obtained with a standard setup. Standard setup has the same configuration, but with an additional intact pump pulse directly coming from the laser, which is synchronized with the SC through an optical delay line. The impact of the excitation strategy on the generation of M-CARS spectra was successfully measured with each system for the vibrational modes at 1296 cm$^{-1}$, 1432 cm$^{-1}$, 2848 cm$^{-1}$, 2874 cm$^{-1}$, 2940 cm$^{-1}$, 2964 cm$^{-1}$, corresponding to $CH_2$ twist, $CH_2$ bend, $CH_2$ symmetrical stretch, $CH_2$ asymmetrical stretch, $CH_3$ symmetrical stretch and $CH_3$ asymmetrical stretch, respectively. A zoom in the area of the C-H bonds between 3050 cm$^{-1}$ and 2700 cm$^{-1}$ (inset of Fig. 5) underlines the weak decrease of peak contrast, induced by spectral broadening of the pump pulse carried by the $LP_{11}$ mode. Indeed, the spectral bandwidth measured at FWHM for the vibrational mode at 2848 cm$^{-1}$ increases from 18 cm$^{-1}$ (when obtained with the standard setup) to 28 cm$^{-1}$ with our self-referenced setup. Thus, the self-referenced setup has a drop of spectral resolution. Note however, that such a drop remains limited, and it does not entail a significant deformation of the spectral shape (so that the $CH_2$ asymmetrical stretching mode could still be clearly identified). Moreover, despite the low spatial overlap between $LP_{01}$ and $LP_{11}$ modes, the obtained CARS signal remains strong enough to be clearly measured. The spatial resolution of our M-CARS imaging setup is limited by the smallest beam profile of either pump, or Stokes wave. Since the beam profile of the SC is close to a Gaussian beam, the spatial resolution remains close to its optical limit.

When observing the spurious signal that emerges above the electronic noise floor of our system, seems to ensure the maximum signal-to-noise ratio possible. To validate this claim, we made a series of CARS measurements by using different monochromatic pump powers. Each value of pump power is strong enough to obtain both the vibration signature and the non-resonant background. We observed that, when the pump power increases, those two signatures grow larger according to the square of the power. The difference of signal amplitudes between these signatures remains a constant when the input pump power is varied (see details in section IX). Therefore, it seems unnecessary to increase the pump power further, when such a non-resonant background is already observed.

It is also important to note that a pump power increase at 1064 nm could be achieved thanks to the addition of a miniaturized Nd:YAG amplifier at the fiber output, placed before the microscope. This allowed us to obtain a higher CARS signal, without affecting the broadband Stokes wave. However, pump amplification remains useful whenever no significant background noise is generated in the CARS spectra. For more details, please refer to the section X.

## VI. Multimodal Imaging

In order to illustrate the ability of our system to produce images using various nonlinear processes, we performed a multimodal imaging of a curcumin superparticle (Fig. 6(a)). The M-CARS spectrum was exploited to observe the inter-ring signature of C=O, and to retrieve the CARS image (Fig. 6(b)). Additionally, the spectrum registered from 525 nm to 575 nm was used to reconstruct second harmonic generation (532 nm) and two-photon fluorescence images (Fig. 6(c)).

Pure curcumin is obtained after purification of rhizomes of *Curcuma longa*, as described in [24]. Then, curcumin, dissolved in acetone, is dropped in a higher volume of water to form slowly curcumin microcrystals, which were obtained after filtration. Curcumin, a polyphenol derivative, shows interesting antioxidant [26], anti-inflammatory [27] and neuroprotective [28] effects. It is important its visualization after cell culture treatment or *in vivo* administration in biological samples, therefore it is important to analyze its biodistribution.

## VII. M-CARS Simulation

Beyond the experimental demonstrations, we also numerically simulated the M-CARS signal of a fictive sample. This calculation was carried out after the spatio-temporal simulation of bimodal propagation in the microstructure fiber, which constitutes a complete simulation of the imaging system including the broadband laser emission.

The complex envelopes $A_0$ and $A_1$ of signals propagating on the $LP_{01}$ and $LP_{11}$ modes, respectively, were filtered in order to drastically attenuate spectral components whose wavelengths are shorter than that of the pump, and to evaluate the response of a typical sample used in CARS spectroscopy. The model used to simulate the sample under test is based on the same equation, where linear dispersion and attenuation have been removed. We choose to consider a sample inspired by Paraffin with a physical length of 1mm, $\tau_1=1/(2\pi\times85.5\text{THz})$, $\tau_2=0.32\text{ps}$. The corresponding Raman response can be visualized in Fig. 7: the resonance is at 85.5 THz, similar to the symmetric stretching of $CH_2$ bond of paraffin.

We developed the code allowing us to simulate the M-CARS process for specific samples and for a given source capable of operating in several temporal domains, which is from the femtosecond to the nanosecond regime. The mastery of the theoretical approach is of paramount importance, in order to properly design the CARS spectroscopy with the desired characteristics.

## VIII. Signal-to-noise Ratio Saturation in M-CARS

By increasing the pump power (at 1064 nm) we observed the growth of the C-H vibrational signature without its spectral shape modification (Fig. 8(a)). The higher the pump power, the higher the CARS signal. To evaluate the signal to nonlinear background noise ratio of this measurement, we normalized the obtained spectra at each pump power, as illustrated in Fig. 8(b). Note that, since the vibrational signature and the non-resonant background both evolve with the square of the pump power, no difference was observed in the signal-to-noise ratio, determined as a ratio between the maximum vibrational peak and the non-resonant background. Increasing the pump power beyond the threshold value for the generation of the non-resonant background is then no longer necessary, as it would only increase the power sent to the sample, up to the point of possibly damaging it.

## IX. M-CARS Improvement by the Pump Amplification

When the pump beam propagates along the MOF, a supercontinuum is generated, while a portion of the wave at 1064 nm serves as a pump for the CARS process. However, the optimal ratio between supercontinuum power density and power at 1064 nm has to be properly fixed, in order to maximize the M-CARS efficiency. Boosting the pump at the fiber output without changing the supercontinuum could be an interesting possibility. Therefore, after MOF we placed a miniature Nd:YAG crystal, pumped by 808 nm diode, in order to amplify the spectral region around 1064 nm of SC. In such a way, we obtained a selective amplification of wavelengths around 1064 nm, which serves as a pump beam for M-CARS signal generation. However, due to the spatiotemporal distortion of the pump beam, the amplification measured at 1064 nm was distributed across time, and only part of the amplified beam remains synchronous with the broadband Stokes beam. Yet, 20% of amplification of the 1064 nm beam resulted in a 20% increase of the M-CARS signal (Fig.9).

## X. Conclusion

In this work, we introduced a novel self-referenced M-CARS setup based on spatial division multiplexing in a microstructured fiber. The quasi-single-mode microstructured fiber permitted the simultaneous propagation of two spatial modes, experiencing different spectral evolutions. The first beam was coupled to the $LP_{01}$ (fundamental) mode, with a power strong enough to obtain a supercontinuum generation with a bandwidth spanning over 1000 nm. Conversely, the second beam was coupled to the $LP_{11}$ mode, with a power low enough to maintain a quasi-monochromatic beam, only weakly distorted by nonlinear effects and cross-phase modulation between the two modes. Such a different spectral broadening behavior was achieved thanks to the low overlap integral between the two fiber modes. Based on the use of the two spatial modes, we could realize a standalone M-CARS system, free from any additional delay line. This system's spectral and spatial resolution remained almost unaffected with respect to a standard CARS system, allowing to clearly identify the vibrational lines of a paraffin sample. We also demonstrated that the light emitted by our microstructured fiber system can be selectively amplified with a miniature Nd:YAG crystal. This approach allowed to point out that the four-wave-mixing process involved in CARS can be governed by a minimum initial power level, which permits to optimize the signal-to-noise ratio. Beyond M-CARS imaging, our system based on spatial multiplexing was successfully applied to multimodal imaging, encompassing parametric multiphoton and second-harmonic fluorescence imaging.

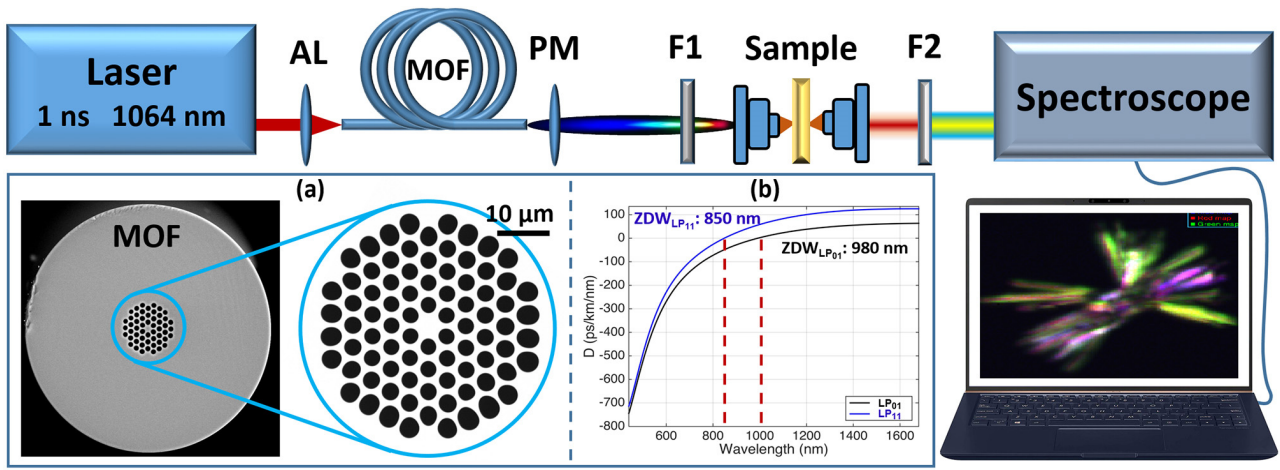

Fig. 1. Schematic of experimental setup for M-CARS experiment. AL, aspherical lens; MOF, microstructured optical fiber; PM, parabolic mirror; F1, long-pass filter at 1000 nm; F2, short-pass filter at 1000nm. Sample is fixed on the 2D scanning stage. (a) Transverse profile of the microstructured fiber, obtained with an electronic microscope. (b) Dispersive curves of the $LP_{01}$ and $LP_{11}$ modes.

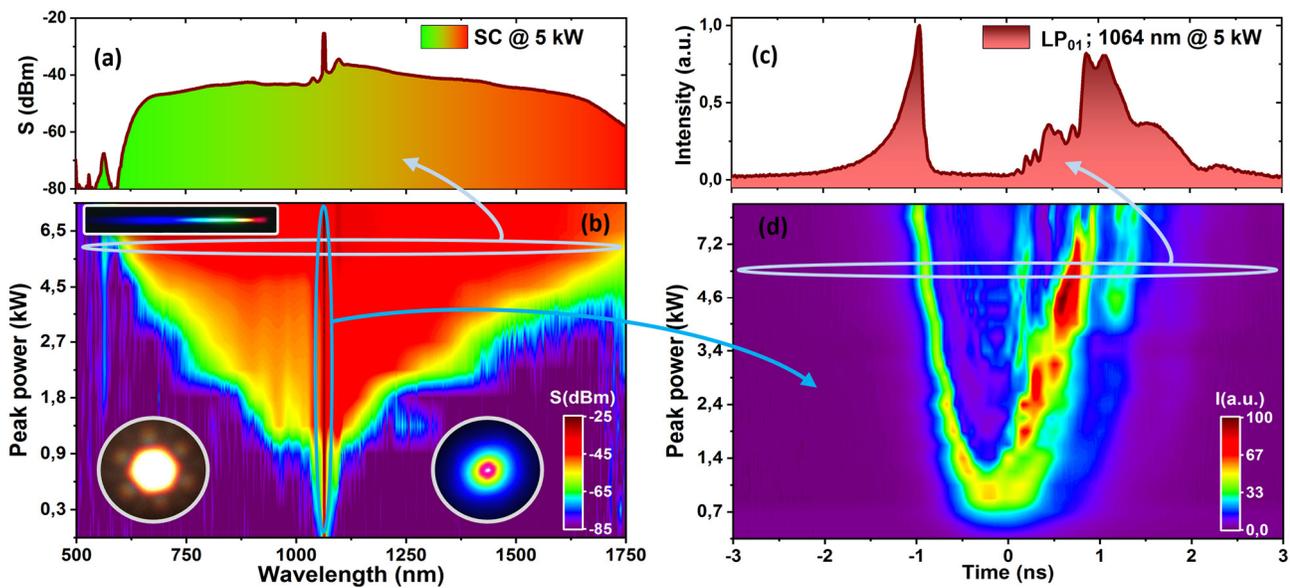

Fig. 2. Beam characteristics at the fiber output. (a) SC spectrum. (b) Spectral evolution of the output spectrum versus input peak power. Insets: output beam profile in the visible (left) and in the near infrared (right) domains. (c) Temporal profile of the pump wave at 1064 nm after nonlinear propagation. (d) Temporal distribution of output energy at 1064 nm versus input peak power. For our M-CARS experiment the peak power injected in the $LP_{01}$: 5 kW, $LP_{11}$: 0.25 kW.

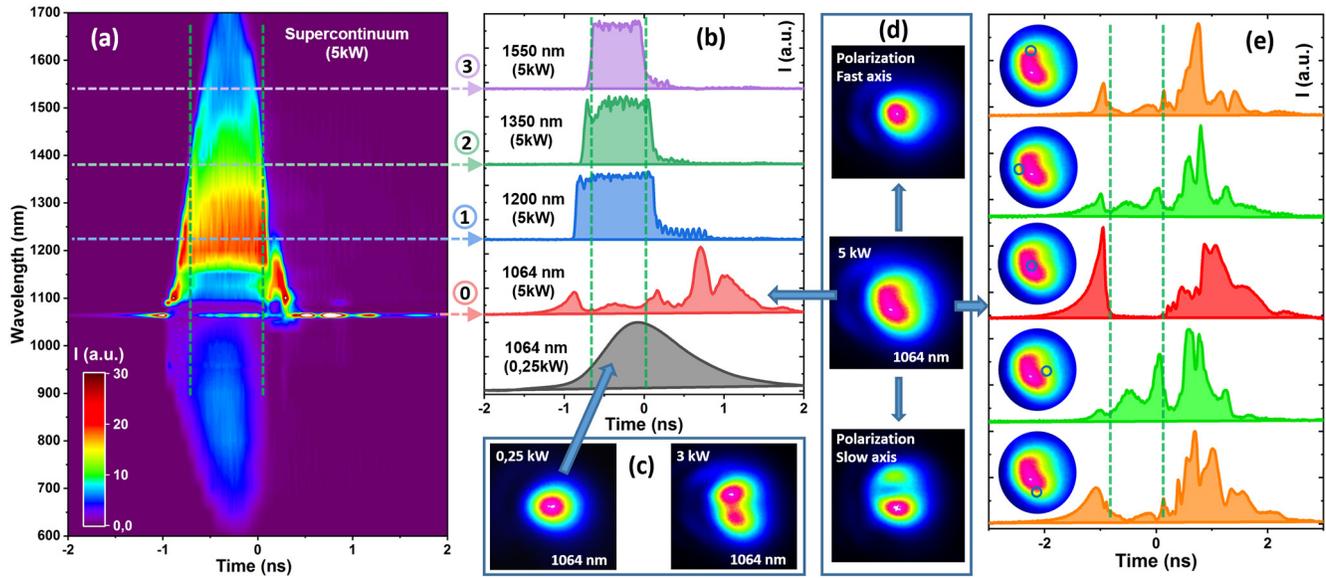

**Fig. 3.** (a) Spectro-temporal energy repartition at the fiber output for 5 kW of input peak power. (b) Temporal profile of the output beam for different wavelengths. Intact pump (low power, linear propagation) also presented. (c) Spatial beam shape of 1064 nm at the fiber output for low and 3 kW input peak powers. (d) Spatial beam shape of 1064 nm at the fiber output for 5 kW of input peak power. The polarization decomposition of this profile illustrates the few-modal nature. (e) Decomposition of the temporal profile of the pump wavelength towards the different parts of the output spatial profile. The range between dashed lines reveals the synchronization of remaining pump and Stokes wavelengths.

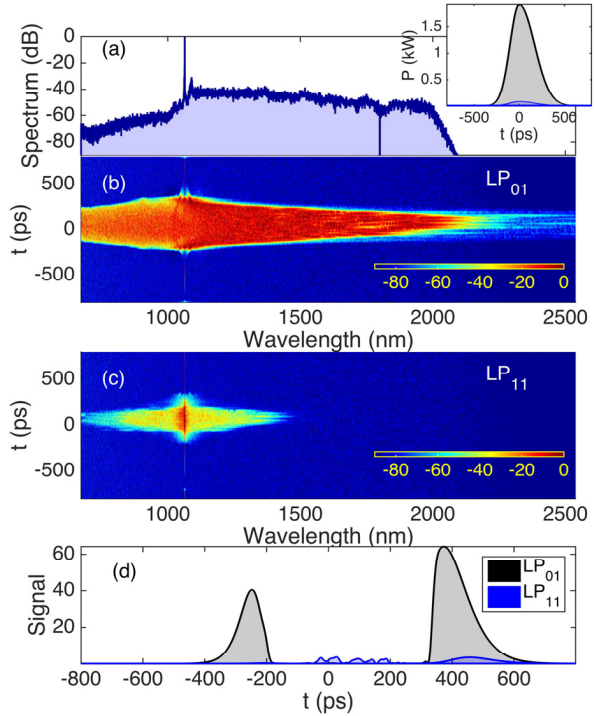

**Fig. 4.** (a) Numerically calculated multimodal spectrum at the output of the MOF. (b) Spectrogram of mode $LP_{01}$. (c) Spectrogram of mode $LP_{11}$. (d) Temporal intensity profiles of the pulse envelopes of the two modes at the fiber output at 1064nm with a filter bandwidth of 25 GHz. Inset: input condition for the two modes.

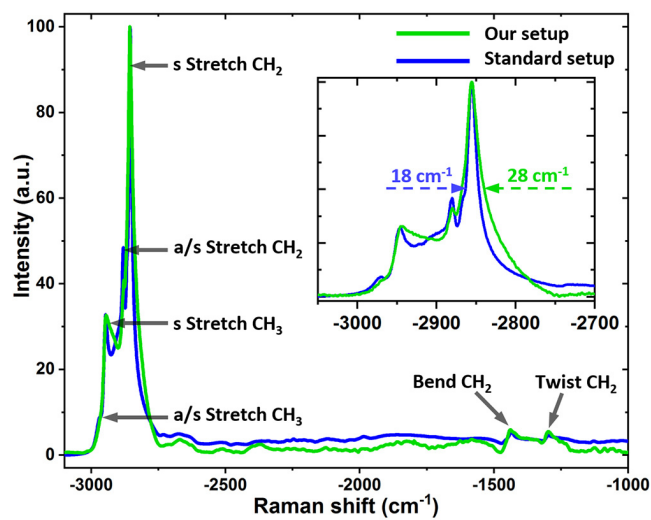

**Fig. 5.** M-CARS signal of paraffin sample obtained with selective coupling (green curve) in microstructure fiber, compared with the signal obtained with a standard M-CARS setup including a delay line synchronization stage (blue curve). Inset: zoom of the C-H zone, revealing a slight drop of spectral resolution with our setup, when compared with the standard setup.

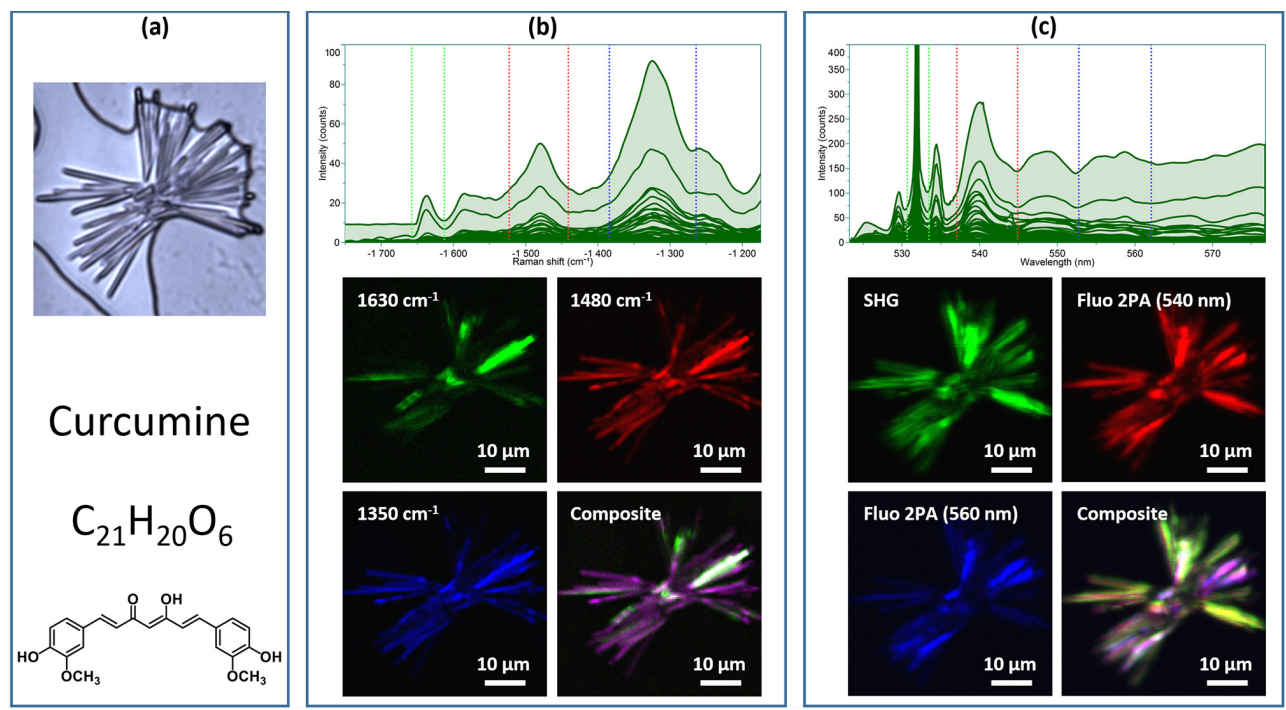

**Fig. 6.** Multimodal imaging of curcumin superparticle. (a) Bright-field microscope image. (b) M-CARS image from three different resonant peaks. (c) Second harmonic generation (SHG) and nonlinear fluorescence (2PA) image.

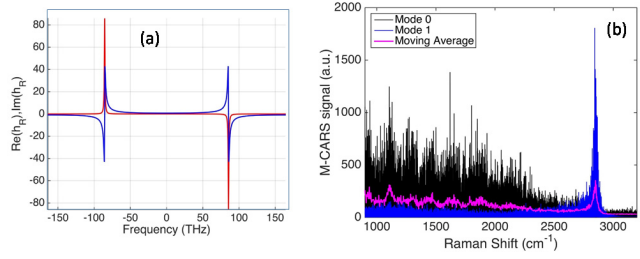

**Fig. 7.** M-CARS Simulation. (a) Representation of vibrational mode at 85.5 THz (2852 CM-1). (b) M-CARS spectrum produced by the four-wave mixing M-CARS process obtained from the bimodal excitation of the microstructured fiber.

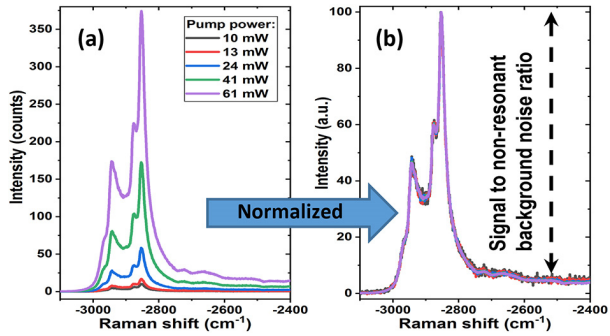

**Fig. 8.** CARS signal to non-resonant noise ratio is invariant. (a) CARS signal of paraffin in the C-H zone, for different input pump (1064 nm) powers. (b) Normalized curves from (a) in order to calculate the signal-to-noise ratio for all different input power.

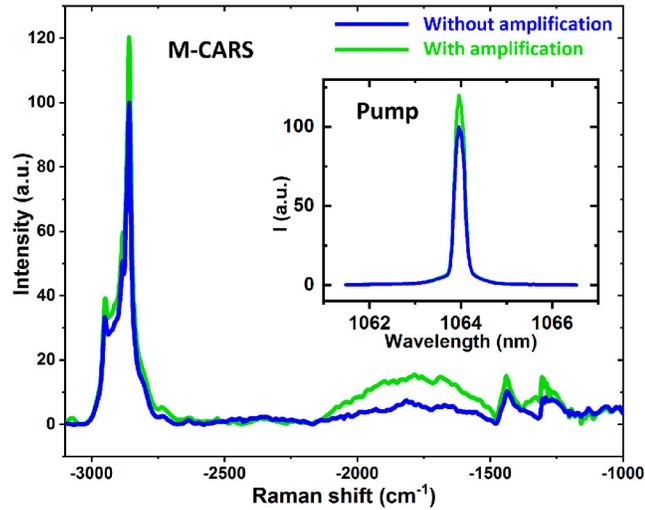

**Fig. 9.** M-CARS signal improvement integrating an amplification step of the pump beam by using Nd:YAG crystal. Comparison of M-CARS signals obtained on a paraffin sample with and without amplification. Inset: amplification of the pump beam @1064 nm.